\newcommand{\Ab}{\mathbf{A}}

\newcommand{\Cb}{\mathbf{C}}

\newcommand{\ab}{\mathbf{a}}
 
\newcommand{\hb}{\mathbf{h}} 
\newcommand{\mb}{\mathbf{m}} 

\newcommand{\nb}{\mathbf{n}}

\newcommand{\ub}{\mathbf{u}}
\newcommand{\vb}{\mathbf{v}}

\newcommand{\Tbm}{\mbox{\bfseries\sffamily{T}}}

\newcommand{\Zbm}{\mbox{\bfseries\sffamily{Z}}}

\newcommand{\abels}{\mbox{\bfseries\sffamily{a}}}
\newcommand{\bels}{\mbox{\bfseries\sffamily{b}}}
\newcommand{\cels}{\mbox{\bfseries\sffamily{c}}}
\newcommand{\dels}{\mbox{\bfseries\sffamily{d}}}

\newcommand{\nex}{\mbox{\bfseries\sffamily{n}}}

\newcommand{\delk}{\mbox{\sffamily{d}}}
\newcommand{\gels}{\mbox{\bfseries\sffamily{g}}}
\newcommand{\kels}{\mbox{\bfseries\sffamily{k}}}

\newcommand{\sels}{\mbox{\bfseries\sffamily{s}}}
\newcommand{\qbels}{\mbox{\bfseries\sffamily{q}}}

\newcommand{\lels}{\mbox{$\log(\nex)$}}

\newcommand{\mub}{\mbox{$\boldsymbol{\mu}$}}
\newcommand{\nib}{\mbox{$\boldsymbol{\nu}$}}

\newcommand{\Ael}{\mathbb{A}}
\newcommand{\Eel}{\mathbb{E}}
\newcommand{\Kel}{\mathbb{K}} 
\newcommand{\Mel}{\mathbb{S}} 
\newcommand{\Nel}{\mathbb{N}} 

\newcommand{\Del}{\mathbb{D}}
\newcommand{\Hel}{\mathbb{H}}
\newcommand{\Gel}{\mathbb{G}}
\newcommand{\Sel}{\mathbb{M}}
\newcommand{\Pel}{\mathbb{P}}

\newcommand{\frakR}{\mbox{\slshape\sffamily{P}}}
\newcommand{\frakK}{\mbox{\slshape\sffamily{K}}}
\newcommand{\frakG}{\mbox{\slshape\sffamily{G}}}
\newcommand{\frakA}{\mbox{\slshape\sffamily{A}}}

\newcommand{\fourA}{\mathcal{A}}

\newcommand{\bzero}{\mathbf{0}}

\newcommand{\Real}{\mathbb{R}}
\newcommand{\Ratio}{\mathbb{Z}}

\newcommand{\caM}{\mathcal M}

\newcommand{\caV}{\mathcal V}
\newcommand{\caD}{\mathcal D}

\newcommand{\caE}{\mathcal E}
\newcommand{\caB}{\mathcal B}

\newcommand{\caR}{\mathcal R}
\newcommand{\caS}{\mathcal S}

\def\dive{\mbox{div\,}}
\def\diag{\mathop{\rm diag}}

\def\tr{\mathop{\rm tr}}

\newcommand{\dif}{\mathrm{d}}

\newcommand{\UNIVPM}{Universit\'a Politecnica delle Marche, 60131 Ancona, Italy}

\documentclass[epj]{svjour}

\usepackage{graphics}
\usepackage{latexsym}
\usepackage{amsfonts,amssymb,amsmath, amsbsy}
\usepackage{epsfig}
\usepackage{epic}

\begin{document}
\title{A Continuum Theory for Scintillating Crystals}
\author{Fabrizio Dav\'{\i}}
\institute{DICEA, \UNIVPM}

\date{Received: date / Revised version: date}

\abstract{
We obtain, by starting from the balance laws of a continuum endowed with a vectorial microstructure and with a suitable thermodynamics, the evolution equation for the excitation carriers in scintillating crystals. These equations, coupled with the heat and electrostatic equations, describe the non-proportional response of a scintillator to incoming ionizing radiations in terms of a Reaction and Diffusion-Drift system. The system of partial differential equations we arrive at allows for an explicit estimate of the decay time, a result which is obtained here for the first time for scintillators. Moreover we show how the two most popular phenomenological models in use, namely the \emph{Kinetic} and \emph{Diffusive} models, can be recovered, amongst many others, as a special case of our model. An example with the available data for NaI:Tl is finally given and discussed to show the dependence of these models on the energy of ionizing radiations.
} 

\maketitle

\section{Introduction}

Scintillating inorganic crystals played a major role in the high-energy physics experiments at CERN as particle collision detectors: indeed they acts as wavelength shifters which convert ionizing radiations into photons and hence, into visible light which in turn can be collected by photomultiplier devices.

A great deal of attention was accordingly dedicated to the many aspects of these crystals, from the better understanding of the basic phenomena and its implications with material engineering and crystal growth, to the energy conversion mechanisms as well as to the analysis of a specific crystal  \cite{LECOQ}, \cite{Birks13}, \cite{ISKO92}, \cite{Anne09}. As far as theoretical and experimental physics are concerned, the paper devoted to the analysis of scintillation in specific materials are countless, \emph{see e.g.\/} the CCC project list of publications \cite{CCC} or the list of references provided in the recent review papers \cite{DAB18}.

A very important and popular research trend in scintillators is concerned with phenomenological models: in these models the scintillation is represented by the means of excitation carrier densities which plays the role of mesoscopic variables which depends on time and evolves according to rate-type equations borrowed from the kinetic of chemical reactions. In the spatially non-homogeneous case terms which account for space diffusion and sometimes drift of the excitation carriers are added to the rate equations. The merits of these models with respect to other approaches are illustrated in some detail into \cite{BM12} (\emph{vid. also\/} \cite{SW15} and the references quoted therein and also the most recent result presented into \cite{VA17}). Indeed scintillation starts at a microscopic scale but the crystals are growth and cut in massive specimen, namely of the order of centimeters and these models well account in simple but non simplistic way for the behavior at a scale which is between the microscopic and the macroscopic one. To this regard it is worth to mention that there are two major computer codes which fills the gap between these two scales, namely GEANT \cite{GEANT} and LiTrAni \cite{LITRANI}: nevertheless phenomenological models are mandatory in order to perform crystal engineering and design based on a restricted set of parameters that can be extracted form experiments.

The aim of this paper is to encompass the various phenomenological models today in use within a thermodynamically consistent phenomenological theory, based on the mechanics of continua with microstructure \cite{CA00}, in order to obtain a very general set of partial differential equations like it was done \emph{e.g.} for semiconductors. The goal is to arrive at a set of equations which depends on a limited number of parameters which can be identified by the means of experiments and which accounts for scintillation at the mesoscopic scale. 
By using such an approach we may give a precise range of validity for the existing model in terms of the material parameters, we may obtain qualitative and quantitative estimates for the solutions and we can formulate models which, instead of being borrowed from the kinetics of chemical reaction, can derived from the physics of scintillation; this paper wishes to be a first step in this direction.

Before to enter into the details of the present work,  let us recall the basic facts about the physics of scintillation which are important in our model formulation.

The scintillation process in inorganic materials is due to the electronic band structure of crystals: a detailed overview of the underlying mechanism can be found \emph{e.g.\/} into \cite{ISKO92}; to summarize,  an incoming ionizing radiation can excite an electron from the valence band to either the conduction band or the exciton band. This leaves an associated hole behind, in the valence band. The pairs electron-hole $(e-h)$ can travel independently or by means of "excitons" $(exc)$ which are loosely bound electron-hole pairs which wander through the crystal lattice. All these three three excited state entities (electrons, holes, and excitons) are referred as \emph{excitation carriers}Ó. Impurities, like dopant centers, create electronic levels in the forbidden gap and the traveling carriers $(e-h)$ and  $(exc)$ can recombine radiatively in the  impurity centers by emitting scintillation light (photons). However some carriers can recombine by "quenching" without generation photons: the two processes are competitive and with a strong non-linear character not yet well understood, as pointed out even in the most recent researches \cite{VA08}-\cite{KD12}.

Scintillation is indeed a very a complex phenomena which happens at different length and time scales: a complete and detailed analysis of the various scales and their relations is presented into \cite{VG14}: there seven length scales related at the various aspects of scintillation within a bulk crystal were described. In the present treatment we find instrumental to lump these scales into three:
\begin{itemize}
\item Microscopic scale: it is the atomic scale of the energy conversion of the ionizing radiation into excitation carriers within the activator nucleus;
\item Mesoscopic scale: it is the scale of the track along which the excitons carriers decays after generating a photon population;
\item Macroscopic scale: it is the scale of the light propagation within the bulk crystal.
\end{itemize}

Here we shall deal only with the first two scales, the interaction between mesoscopic and macroscopic scale being described only in terms of the temperature: the model we obtain here will be also the starting point for a correct mathematical description of two of the most important design parameter for scintillating crystals, namely the \emph{Light yield}, namely the ratio between the energy converted into photons and that of the ionizing radiation, and of  the \emph{Decay time}, namely  how long it takes the excited states to de-excite and give off light. These two aspects, as well as
the study of the interactions with other relevant macroscopic variables like deformations, electromagnetic fields and crystal defects will be the object of forthcoming papers \cite{DA171}, \cite{DA172}.

We remark that the physic of scintillation reminds that of semiconductors and indeed the results we obtain are similar to those obtained for semiconductors into \cite{AGH02}: moreover in the mathematical treatment of our model we follow closely \cite{MIE11}; however there are many differences between the two phenomenologies, as pointed out into \cite{JAJP07}, and it is important to remark that our results follow from the balance law of the mechanics of micro-structured continua and the associated thermodynamics.

The paper is organized as follows: in \S.2 we deal with the basics of excitation carriers creation at the microscopic scale and we define, by the means of an approximate solution of the Bethe-Bloch equation \cite{ULMA10}, the dependence of these excitation carriers on the initial energy and on the material parameters. By a magnification/scaling procedure then we find which part of these excitations are "visible" at the mesoscopic scale and at the beginning of \S.3 we define in terms of this part the basic state variable of our model, the $M-$dimensional \emph{Excitation density vector} $\nex$.

In \S.3, which is the main part of this paper, we postulate a microforce evolution equation which represents, from a physical point of view, the continuity equation for the electric current which is associated to an electrostatic and internal (\emph{i.e.\/} dependent on the absolute temperature $\theta$) energy within a suitable control volume $\caR$ were we assume the conversion into photon is confined within. By the means of a procedure which is standard in continuum mechanics, from the balance of energy and the entropy inequality written at the mesoscopic scale, then we arrive at a fairly general reaction-diffusion equation with Neumann boundary conditions:
\begin{equation}\label{prima}
\dive(\Mel(\nex\,,\theta)[\nabla\mub])-\Hel(\nex\,,\theta)\mub=\dot{\nex}
\end{equation}
where $\mub=\mub(\nex)$ is the \emph{Scintillation potential}, the derivative of the Gibbs free-energy with respect to $\nex$, and $\Mel$ and $\Hel$ are two semi-positive definite $M\times M$ matrices. Such a reaction diffusion equation is coupled with the Laplace equation for electrostatic at the mesoscopic scale and with the heat equation at the macroscopic scale, as in \cite{AGH02} for semiconductors. As it is shown into \cite{MIE11}, equation (\ref{prima}) can be put in an equivalent gradient-flow formulation which involves a conjugate dissipation potential $\psi^{*}(\nb\,,\mub)$ which was previously obtained from the entropy inequality:
\begin{equation}\label{seconda}
\dot{\nb}=-D\psi^{*}(\nb\,,\mub)\,,
\end{equation}
where $D$ denotes the Frech\'{e}t derivative. In the last section, \S.4, we specialize (\ref{prima}) from a constitutive point of view by expressing $\Mel$ in terms of a \emph{Mobility matrix} to arrive at a reaction diffusion-drift equation in terms of the variable $\nex$ which recovers, within our treatment, the model of \cite{VA08} (\emph{vid.\/} also \cite{WGLU11}-\cite{LU17}): by a suitable adimensionalization then we show also that the two most popular phenomenological model, the \emph{Kinetic} \cite{BM09}, and the \emph{Diffusive} one \cite{BM12} can be obtained as special cases of (\ref{prima}). 

As far as the notation is concerned, we shall use as long as possible the intrinsic notation. In particular, let $\caE$ be the three-dimensional point space and $\caV$ its associated vector space, we denote with boldface lower case the elements of $\caV$, \emph{i.e.\/} the three-dimensional vectors $\ub\in\caV$; we denote with boldface uppercase the second order tensors $\Ab$ which maps $\caV$ into itself, \emph{i.e.\/} $\Ab\ub=\vb$, $\forall\,\ub\,,\vb\in\caV$.  We denote with boldface sans-serif lower-case letter (as $\nex$), the M-dimensional vectors in $\Real^{M}$, whereas boldface sans-serif upper-case letters (like \emph{e.g.} $\Tbm$) will denote $M\times 3$ matrices. We shall also make use of $M\times M$ matrices like \emph{e.g.\/} $\Del$: other higher-order matrices shall be defined when necessary.

\section{From microscopic to mesoscopic scale: the excitation density}

\subsection{The excitation carriers and the energy balance on a track}

When an Ionizing Radiation ($X-$, $\gamma-$, $\alpha-$ or $\beta$-ray for instance) interacts with a scintillator, its energy $E^{*}$ creates excited-state \emph{electron} and \emph{holes} $(e-h)$ which transport the energy through the material: the energy can be also transported by \emph{excitons} $(exc)$ which are bounded electron-holes pairs: all these excited states are collectively referred as \emph{excitation carriers}. According to \cite{JA07}, these excitation carriers travel for about 10 micron within the material with a path characterized by many kinks and bends: however in the initial stages of the scintillation phenomena, these excitation carriers travels in a cylindrical region whose length is few nanometers and whose radius is about $2\div 3$ nanometers. In this section we deal with the initial stages of scintillation phenomena in this cylindrical region whose length $L$ is called the \emph{mean free-path} .

Let $x$ be the point where the energy $E^{*}$ hits the scintillator: we define the \emph{Excitation track} at $x$ the cylindrical region $\Omega\equiv\caS\times L$ where $\caS$ is the disk of radius $r$ centered at $x$ and $L$ is the mean free-path whose \emph{track coordinate} is $z\in(0\,,L)$ (Fig. 1).

\begin{center}
\begin{picture}(200,150)
\put(0,-80){\scalebox{0.90}{\includegraphics[width=\linewidth]{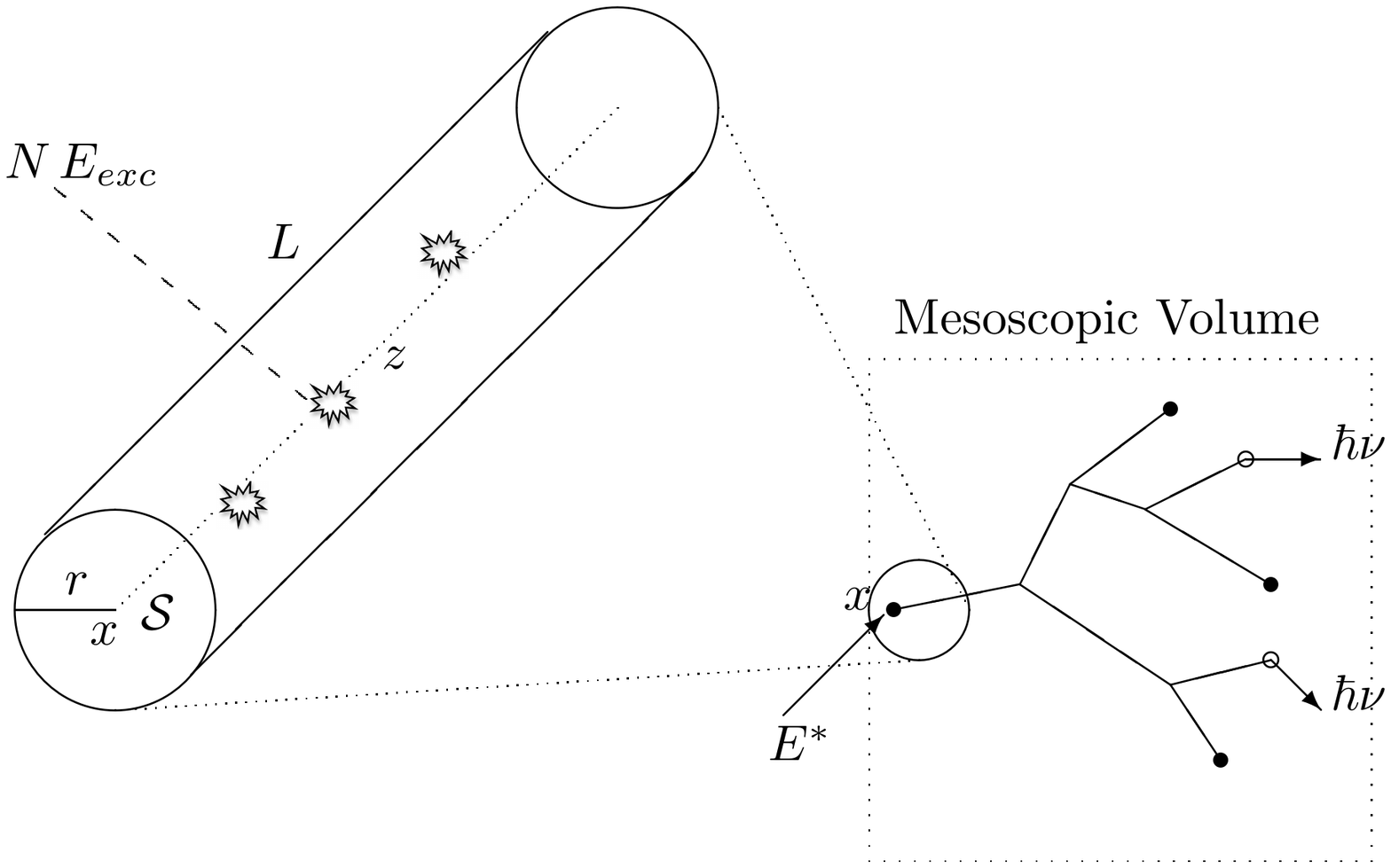}}}

\end{picture}

Fig. 1 - A schematic of the excitation track $\Omega$.
\end{center}

The total energy $W$ of the excitation carriers contained in the track is given by:
\begin{equation}\label{energy1}
W=\int_{\Omega}N\,E_{exc}\,\dif\Omega\,,\quad E_{exc}=\beta E_{gap}\,,
\end{equation}
where $N\geq 0$ is the density of the excitation carriers in $\Omega$, whose dimension is m$^{-3}$,  $E_{exc}$ denotes the \emph{excitation energy}, $E_{gap}$ is the forbidden \emph{gap energy} between the valence and the conduction bands in the atom and the parameter $\beta$, (whose lower limit is $\beta=2.3$ \cite{LWB93}) depends on the specific scintillator. We remark that both $\beta$ and $E_{gap}$ are two constitutive parameters which characterize the scintillator.

The track radius $r$ and the mean free-path length $L$ depend on the energy $E^{*}$: moreover, the dependence of the excitation carriers on such energy is a strong non-linear one and such a non-proportionality is, according to \cite{ROD97}, "an intrinsic property of the crystal".  As far as the dependence on the energy of the radius, we follow \cite{JA07} on assuming that $r$ is constant, or rather, that its variation is an order of magnitude lower than the corresponding variation of $L$.

Let $E_{0}=E(z=0)=E^{*}$  and $E_{L}=E(z=L)$, then, since we assume that all the excitation carriers are contained within the track, the energy balance on $\Omega$ reads:
\begin{equation}\label{energy2}
E_{0}=W+E_{L}\,;
\end{equation}
if we define the \emph{Average number of excitation carriers} on the track section $\caS$ as:
\begin{equation}\label{average}
\bar{N}=\frac{1}{\pi\,r^{2}}\int_{\caS}N\,\dif\caS\,,
\end{equation}
then (\ref{energy1}) can be rewritten in terms of (\ref{average}), upon the assumption that $E_{exc}$ be constant, as:
\begin{equation}\label{WmeanN}
W=\pi\,r^{2}\,E_{exc}\int_{0}^{L}\bar{N}\,\dif z\,,
\end{equation}
and accordingly, from (\ref{average}) and (\ref{WmeanN}) we have:
\begin{equation}\label{microscopicbalance}
W=E_{0}-E_{L}=-\int_{0}^{L}\frac{dE}{dz}\,\dif z=\pi\,r^{2}\,E_{exc}\int_{0}^{L}\bar{N}\,\dif z\,.
\end{equation}

The track energy balance (\ref{microscopicbalance}) can be localized to arrive at the relation between $\bar{N}$ and the rate of change of the energy along the track:
\begin{equation}\label{bethe1}
\bar{N}=-\frac{1}{\pi\,r^{2}\,E_{exc}}\frac{dE}{dz}\,,
\end{equation}
which is indeed the differential formulation of equation (1) of \cite{JA07}. 

The dependence of $\bar{N}$ and of the track length $L$ on the initial energy $E^{*}$ can be evaluated by the means of the Bethe-Bloch equation \cite{SE52}
\begin{equation}
S(E)=-\frac{dE}{dz}\,,
\end{equation}
where $S(E)$ represents the \emph{Stopping Power} which characterizes the material.

The term $S(E)$ has many different formulations, depending on various correction terms which account for different ranges of validity of the Bethe-Bloch formula (\emph{vid. e.g.\/} \cite{ZIE99} and \cite{INOKU71}); some of them were used to get either approximate analytical expressions as in \cite{SW15}, \cite{JA07}, \cite{LERA09}, or numerical evaluation, as in \cite{VA08}, \cite{BMSVW09}, \cite{BM10},  of the dependence of the track length $L$ on $E^{*}$.

In the present treatment, since we are interested into an analytical expression for both $\bar{N}$ and $L$ in terms of $E^{*}$,  we follow \cite{ULMA10} (\emph {vid. also} \cite{ULM07}, \cite{ULSC11}), which gives an approximated solution for the Bethe-Bloch equation which  accounts for various correction terms (unlike \emph{e.g.\/} in \cite{JA07}) and yields the following expression for $L=L(E^{*})$:
\begin{equation}\label{tracklength}
L(E^{*})=aE^{*}(1+\sum_{k=1}^{\infty}b_{k}(1-e^{-\xi_{k}E^{*}}))\,,
\end{equation}
where the terms $a$, $b_{k}$ and $\xi_{k}$ depends on the material parameters which appears into the expression of the stopping power.\footnote{These parameters are, besides the correction terms,  the \emph{atomic number} $Z$, the \emph{atomic weight} $A$ and the material density $\rho$.}  As it is shown in detail into \cite{ULMA10}, relation (\ref{tracklength}) can inverted into 
\begin{equation}\label{initialenergy}
E^{*}=L\sum_{k=1}^{\infty}c_{k}\,e^{-\lambda_{k}L}\,.
\end{equation}
The parameters in (\ref{tracklength}) and the corresponding ones into (\ref{initialenergy}), can be obtained from those in the Bethe-Bloch equation, as it was done into \cite{ULMA10} for water and other materials of biological interest. However a complete derivation of these parameters for inorganic scintillators is out of the scope of this paper: instead relation (\ref{initialenergy}) can be used for instance to extract $(c_{k}\,,\lambda_{k})$ from $(E^{*}\,,L(E^{*}))$ graphs obtained either experimentally or by numerical integration. As an example a graph $(\log E^{*}\,,\log L(E^{*}))$ obtained in \cite{BM10} is shown in Fig. 2. 

From (\ref{initialenergy}) the energy profile along the track can be obtained as
\begin{equation}\label{trackenergy}
E(z)=(L-z)\sum_{k=1}^{\infty}c_{k}\,e^{-\lambda_{k}(L-z)}\,,
\end{equation}
which by (\ref{bethe1}) yields in turn the excitation density profile along the track:
\begin{equation}\label{trackdensity}
\bar{N}(z)=\frac{1}{\pi\,r^{2}\,E_{exc}}\sum_{k=1}^{\infty}(1-\lambda_{k}(L-z))c_{k}e^{-\lambda_{k}(L-z)}\,.
\end{equation}

We take relations (\ref{tracklength})-(\ref{trackdensity}) as the starting point for the definition of the state variable which shall accounts, at the mesoscopic scale, for the excitation density.

\begin{center}
\begin{picture}(200,180)
\put(0,10){\scalebox{0.85}{\includegraphics[width=\linewidth]{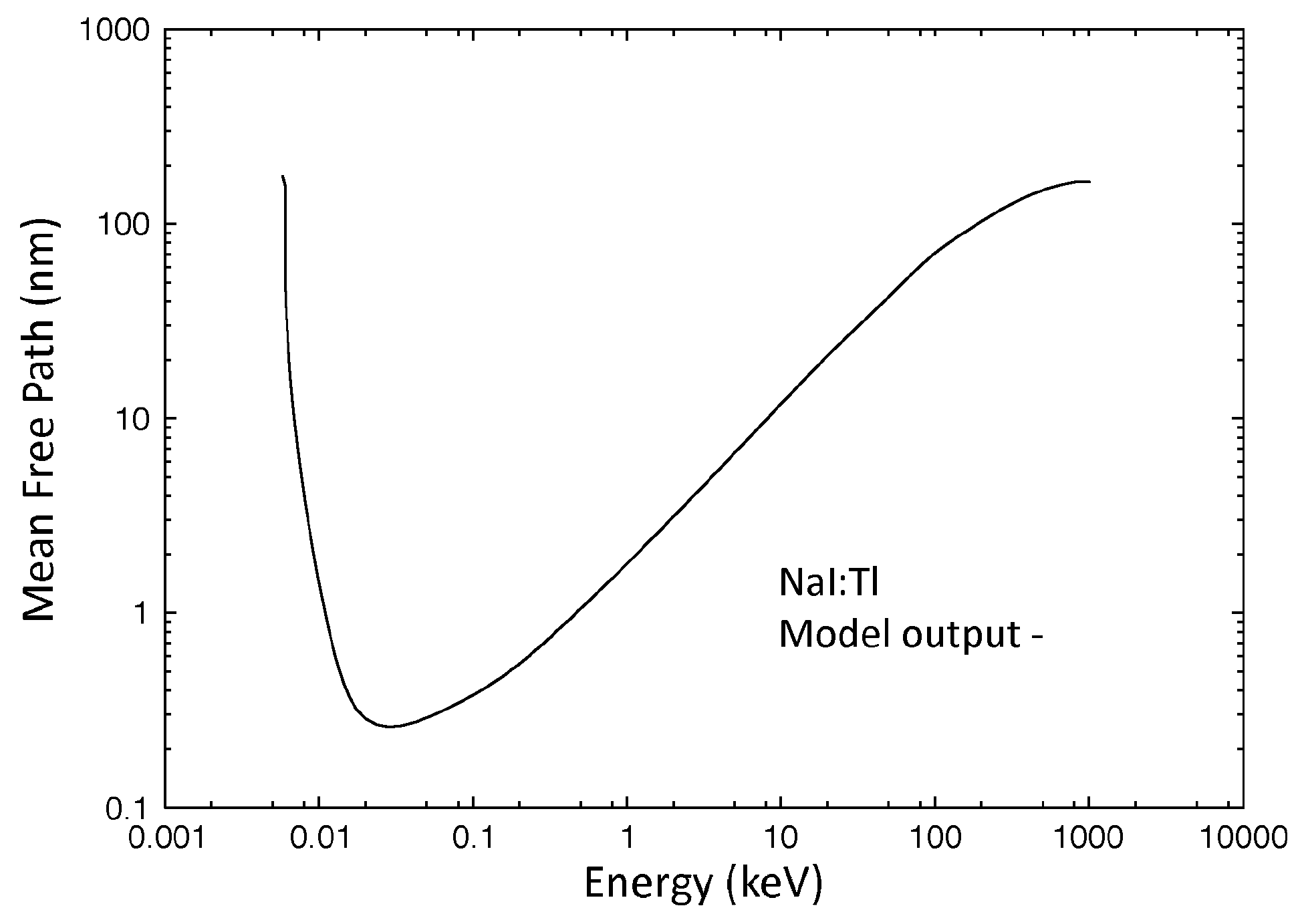}}}
\end{picture}

Fig. 2 - Numerical evaluation of the track length dependence on $E^{*}$ for NaI:Tl,  from \cite{BM10}.
\end{center}

\subsection{Upward to the mesoscopic scale}

Relation (\ref{trackdensity}) describes the distribution of our measure $\bar{N}$ of excitation density along a track: what we need now is to change the scale from the microscopic to the mesoscopic in order the describe the interaction processes which, at this scale, generate the photon emission.

Such a change of scale is done by introducing a suitable \emph{magnifier}, that is a scalar parameter $\sigma\in[0\,,1]$ such that the track dimensions can be rewritten as follows:
\begin{equation}\label{scaleddimension}
r(\sigma)=\sigma\,r\,,\quad L(\sigma)=\sigma\,L\,;
\end{equation}
accordingly, when $\sigma=1$ we have the maximum magnification and hence such value corresponds to the microscopic scale, whereas the transition to the mesoscopic scale is obtained for $\sigma\rightarrow 0^{+}$: we remark that, since:
\begin{equation}
\mbox{vol}\,\Omega(\sigma)=\sigma^{3}\mbox{vol}\,\Omega\,,
\end{equation}
in the limit for $\sigma\rightarrow 0^{+}$ the volume vanishes and the track reduces to the point $x$.

We rewrite (\ref{trackdensity}) in terms of the magnified dimensions (\ref{scaleddimension}) and of the dimensionless track variable $s=z/L(\sigma)$: 
\begin{equation}\label{trackdensity1}
\bar{N}(s\,,\sigma)=\frac{1}{\pi\,\sigma^{2}r^{2}\,E_{exc}}\sum_{k=1}^{\infty}(1-\sigma L\lambda_{k}(1-s))c_{k}e^{-\sigma\lambda_{k}L(1-s)}\,;
\end{equation}
then, by taking the Taylor expansion of (\ref{trackdensity1}) about $\sigma=0$ we obtain
\begin{eqnarray}\label{trackdensity2}
\bar{N}(s\,,\sigma)&=&\frac{1}{\pi\,r^{2}\,E_{exc}}\sum_{k=1}^{\infty}\frac{1}{\sigma^{2}}c_{k}+\frac{1}{\sigma}2c_{k}\lambda_{k}L(1-s)\nonumber\\
&+&\frac{3}{2}c_{k}(\lambda_{k}L(1-s))^{2}+O_{k}(\sigma)\,.
\end{eqnarray}

Since (\ref{trackdensity2}) blows up to infinity for $\sigma\rightarrow 0^{+}$, we must search for a suitable rescaled excitation density whose limit at the mesoscopic scale remains finite. Trivially, such rescaling is:
\begin{equation}
\quad N_{\sigma}(s)=\sigma^{2}\bar{N}(s\,,\sigma)\,;
\end{equation}
looking at (\ref{average}) such a rescaling means that it is the total number of excitation carriers across the track section $\caS$ the physical quantity which remains invariant under the passage to the mesoscopic scale. Accordingly we define
\begin{equation}
 N_{mes}=\lim_{\sigma\rightarrow 0^{+}}N_{\sigma}=\frac{1}{\pi\,r^{2}\,E_{exc}}\sum_{k=1}^{\infty}c_{k}\,,\quad N_{mes}> 0\,,
\end{equation}
which is independent on both the magnifier $\sigma$ and on the normalized track coordinate $s$; it depends on the ionization energy $E^{*}$ which hits the crystal at a point $x$ and at a given time $t$ only by the means of the parameters $c_{k}$.

\section{The evolution equations for the excitation carriers}

\subsection{State variable, balance laws and order parameter}

The scaled excitation carrier density $N_{mes}$ we defined in the previous section  represents, at a point $x$ and at a given time $t$,  a mesoscopic measure of the excitation carriers induced at the microscopic scale by the incoming ionization energy $E^{*}$ which hits the body at $(x\,,t)$.  These excitation carriers however have different possible interaction mechanisms which lead to various kind of recombination (with or without photon creation), annihilation or further carriers generations. Accordingly we can assume that $N_{mes}$ can be represented as the sum of different kinds of excitation carriers, each one associated with a different interaction mechanism:
\begin{equation}
N_{mes}(x\,,t)=\sum_{k=1}^{M}n_{k}(x\,,t)\,,\quad n_{k}\geq 0\,;
\end{equation}
clearly, the greater is $M$, the finer will be the description of the interaction mechanism:\footnote{For instance the models developed into \cite{BM09}, \cite{LGW11}, \cite{SI11},  have $M=2$ (electron-hole and exciton carriers densities), those in \cite{WGLU11} and \cite{BM12} split between the electron and holes with $M=3$, the model developed into \cite{GBD15} has $M=7$ whereas those in \cite{VA08} and \cite{BM10} have unspecified $M$ ($M\geq 11$ in \cite{VA08}).} then we define the state variable which describes scintillation as the \emph{vector of excitation carriers}  $\nex(x\,,t)\in\caM\equiv [0\,,+\infty)^{M}$:
\begin{equation}
\nex(x\,,t)\equiv(n_{1}(x\,,t)\,,n_{2}(x\,,t)\,,\ldots n_{M}(x\,,t))\,.
\end{equation}
We find useful to introduce the $M\times M$ diagonal matrix $\Nel(\nex)\equiv\diag(n_{1}\,,n_{2}\,,\ldots\,,n_{M})$ in such a way we can write, \emph{e.g.\/}, $N_{mes}=\tr\Nel$; we shall also make use of the diagonal matrix $\Nel(\nex)^{-1}$ whose entries are defined by $n_{k}^{-1}$ for $n_{k}\neq 0$ and $0$ for $n_{k}=0$.

Let $\caB$ the three-dimensional region comprised of the crystal and let $\caR\subset\caB$ be the subregion about $(x\,,t)$ where the photons are produced by means of the various recombination processes; to arrive at an evolution equation we start by postulating a general balance law in terms of micro-mechanical quantities \cite{CA00}:
\begin{eqnarray}\label{currentbalance}
&\dive\Tbm-\kels+\bels=\dot{\nex}\,\mbox{   in }\caR\times[0\,,\tau)\,,\\
&\Tbm\mb=\sels\,\mbox{   on }\partial\caR\times[0\,,\tau)\,,\nonumber
\end{eqnarray}
where $\mb$ denotes the outward unit normal to $\partial\caR$: for the sake of notation we shall henceforth set $\caR_{t}\equiv\caR\times[0\,,\tau)$ and $\partial\caR_{t}\equiv\partial\caR\times[0\,,\tau)$. Moreover we assume that on $\caR_{t}$ and $\partial\caR_{t}$ an appropriate measure can be well-defined in such a way that from here to now on we can dispense of the integration variable in the volume and surface integrals.

From a physical point of view such an equation expresses the conservation of charge density  (normalized with respect to the electron charge $e\approx 1.6\cdot10^{-19} C$) within $\caR$ provided the following  identification for the micromechanical quantities is done:
\begin{itemize}
\item $\Tbm(x\,,t)$, the \emph{microstress}, a tensor field which represents the electric current density. Its dimensions are $[\Tbm]=m^{-2}sec^{-1}$: we remark that it is indeed an electric current density normalized w.r.t. $e$;
\item $\kels(x\,,t)$, the \emph{interactive microforce} a vector field which represents the rate of change of charge density generated by excitation carriers: $[\kels]=m^{-3}sec^{-1}$;
\item $\bels(x\,,t)$, the \emph{volume microforce},  a vector field which represents the rate of change of charge density supplied by external sources: $[\bels]=m^{-3}sec^{-1}$;
\item $\sels(x\,,t)$, \emph{surface microforce} a vector field which represents the electric current density which flows through the boundary of $\caR$: $[\sels]=m^{-2}sec^{-1}$;  
\end{itemize} 
it is important to remark that $(\bels\,,\sels)$ represents a system of external actions for $\caR$ in the sense that we may assume that in an experiment they can be controlled and be disposed of. 

The theory developed into \cite{CA00} (\emph{vid. also} \cite{CA000} and \cite{CV90}) requires the definition of a suitable \emph{order parameter} $\dels(x\,,t)$ whose time derivative $\dels\,\dot{}$ expends mechanical power with the micro-mechanical quantities. 

Given the specific physical meaning of the balance law (\ref{currentbalance}) then the components of $\dels\,\dot{}$ must have the dimension of an energy (\emph{e.g.} $[\dot{\delk}_{k}\,]=eV$) and we introduce accordingly a \emph{scintillation potential} $\mub(x\,,t)$ such that:
\begin{equation}
\dels\,\dot{}=\mub\,.
\end{equation}

In the sequel it will appear clear that such a scintillation potential plays the same role of the electrochemical potential in semiconductors: however we prefer not to call it "electrochemical" since neither in scintillators and nor in semiconductors there are chemical reactions at the basis of the relevant physical phenomena indeed. As far as the order parameter is concerned, it represents what in classical physics is called the \emph{action}:
\begin{equation}
\dels(x\,,t)=\int_{0}^{t}\mub(x\,,\tau)d\tau\,.
\end{equation}

\subsection{Thermodynamics}

We begin by assuming that in $\caR$ there exists an \emph{microenergy density}\footnote{The use of the suffix "micro" refers note to the microscopic scale but rather to the micro-mechanical description of the phenomena. The terminology, as well as many underlying ideas are borrowed from (\cite{PPG06}) where the Allen-Cahn and Cahn-Hillard equations are developed within a micro-mechanical theory like the one we are using here (\emph{vid. also } \cite{GU96}).} $\varepsilon=\varepsilon(x\,,t)$ which depends on the mesoscopical variable $\nex(x\,,t)$ and on a set of macroscopical variables to account for the interaction between the "bulk" crystal and the photon production, like it was done into \cite{XB08} for semiconducting ferroelectrics.

A this stage, to make things easier, we limit such a set to the \emph{absolute temperature} $\theta$ in such a way that our model shall depict the scintillation in a non isothermal, rigid and defectless crystal.  The extension to other relevant macroscopic variables, in particular deformation and defects, will be the object of a forthcoming paper \cite{DA171}.

As we pointed out in the description of the physics of scintillation, the excitation carriers induces a charge distribution within the control volume $\caR$ which in turn induces an electric potential $\varphi=\varphi(x\,,t)$ which obeys:
\begin{eqnarray}\label{laplace}
-\epsilon_{o}\Delta\varphi=e\qbels\cdot\nex+q^{*}\,,\mbox{ in }\caR_{t}\,,\nonumber\\
\\
\nabla\varphi\cdot\mb=0\,,\mbox{ on }\partial\caR_{t}\,;\nonumber
\end{eqnarray}
where $\qbels=(q_{1}\,,q_{2}\,,\ldots q_{M})$, $q_{k}\in\Ratio$ is the \emph{charge vector} and $q^{*}$ is the external charge associated with the ionizing radiation. For $\gamma-$ and $X-$rays we have $q^{*}=0$, whereas for $\alpha-$rays it is $q^{*}>0$ and $q^{*}<0$ for $\beta-$rays. We also remark that the Neumann boundary condition (\ref{laplace})$_{2}$ can be satisfied by an appropriate choice of $\caR$. 

Accordingly the energy density can be split into an \emph{electrostatic} and an \emph{internal} energy densities:
\begin{equation}
\varepsilon(\varphi\,,\theta)=\frac{1}{2}\epsilon_{0}\|\nabla\varphi\|^{2}+u(\theta)\,;
\end{equation}
it  easy to show that from the divergence theorem and the boundary condition (\ref{laplace})$_{2}$ we can obtain:
\begin{equation}\label{reducedenergy1}
\varepsilon(\nex\,,\theta)=e\varphi\,\qbels\cdot\nex+u(\theta)\,.
\end{equation}

We further assume that in $\caR$ there exists an \emph{microentropy density} $\eta$ and that like the energy density it can be split into a part which depends on $\nex$ and a part which depends on $\theta$:
\begin{equation}
\eta(\nex\,,\theta)=\eta_{e}(\nex)+\eta_{u}(\theta)\,.
\end{equation}

We write the \emph{balance of microenergy} as
\begin{equation}\label{energybalance1}
\frac{d}{dt}\int_{\caR}\varepsilon=-\int_{\partial\caR}\hb\cdot\mb+\int_{\caR}r+w(\mub)\,,
\end{equation}
where $\hb$ represent the \emph{heat flux}, $r$ the \emph{heat source} and the \emph{microworking} $w(\mub)$, a linear functional of the order parameter,  represents the power expended by the system of external actions $(\bels\,,\sels)$:
\begin{equation}
w(\mub)=\int_{\partial\caR}\sels\cdot\mub+\int_{\caR}\bels\cdot\mub\,.
\end{equation}
By (\ref{currentbalance}) and the divergence theorem then (\ref{energybalance1}) can be localized into:
\begin{equation}\label{localbalance1}
\dot{\varepsilon}=-\dive\hb+r+\Tbm\cdot\nabla\mub+(\kels+\dot{\nex})\cdot\mub\,.
\end{equation}

From the \emph{microentropy inequality}
\begin{equation}
\frac{d}{dt}\int_{\caR}\eta\geq-\int_{\partial\caR}\theta^{-1}\hb\cdot\mb+\int_{\caR}\theta^{-1}r\,,
\end{equation}
we get instead the following local form:
\begin{equation}\label{localentropy}
\dot{\eta}\geq-\theta^{-1}(\dive\hb-r)+\theta^{-2}\hb\cdot\nabla\theta\,.
\end{equation}

When, as it is customary, we introduce the \emph{Gibbs free microenergy}
\begin{equation}
\psi=\varepsilon-\theta\eta\,,
\end{equation}
then from (\ref{localbalance1}) and (\ref{localentropy}) we arrive at the \emph{reduced dissipation inequality}:
\begin{equation}\label{reduceddissipation}
\dot{\psi}+\eta\dot{\theta}-\mub\cdot\dot{\nex}-\Tbm\cdot\nabla\mub-\kels\cdot\mub+\theta^{-1}\hb\cdot\nabla\theta\leq 0\,,
\end{equation}
which will be the starting point for our constitutive prescriptions. We made the following constitutive assumption on the Gibbs free microenergy:
\begin{equation}
\psi=\hat{\psi}(\nex\,,\theta\,,\nabla\theta\,,\mub\,,\nabla\mub)\,,
\end{equation}
and then we require that it will be consistent with (\ref{reduceddissipation}) for all processes; by an argument which is standard in continuum mechanics, then we get that the material response function $\hat{\psi}$, in order to be consistent with (\ref{reduceddissipation}), must obey:
\begin{eqnarray}\label{const1}
\psi&=&\hat{\psi}(\nex\,,\theta)\,;\nonumber\\
\eta&=&-\hat{\psi}_{\theta}(\nex\,,\theta)\,;\\ 
\mub&=&\hat{\psi}_{\nex}(\nex\,,\theta)\,.\nonumber
\end{eqnarray}
Moreover the heat flux must be such that:
\begin{equation}\label{const2}
\hb(\nex\,,\theta\,,\nabla\theta)=-\Cb(\nex\,,\theta)[\nabla\theta]\,,
\end{equation}
with the \emph{Conductivity} $\Cb$ a symmetric and positive-definite second-order tensor, whereas the microstress tensor and the interactive microforce must obey:
\begin{eqnarray}\label{const3}
\Tbm(\nex\,,\theta\,,\nabla\mub)&=&\Mel(\nex\,,\theta)[\nabla\mub]\,,\\
\kels(\nex\,,\theta\,,\mub)&=&\Hel(\nex\,,\theta)\mub\,,\nonumber
\end{eqnarray}
with $\Mel$ and $\Hel$ two symmetric and positive semi-definite $M\times M$ matrices. As a consequence of (\ref{const1}), (\ref{const2}) and (\ref{const3}) we can write the rate of a change of the microenergy density as:
\begin{equation}\label{reducedmicroen}
\dot{\varepsilon}=\theta\dot{\eta}+\mub\cdot\dot{\nex}\,.
\end{equation}

\subsection{Coupled evolution equations}

When we put (\ref{const2}) and (\ref{const3}) into (\ref{currentbalance}), (\ref{const1}) and (\ref{reducedmicroen}) into (\ref{localbalance1}), by taking into account  (\ref{laplace}), then  we arrive at the coupled evolution boundary-value problem for $(\nex\,,\theta)$:
\begin{eqnarray}\label{evolution1}
&\dive(\Mel(\nex\,,\theta)[\nabla\mub])-\Hel(\nex\,,\theta)\mub+\bels=\dot{\nex}\,,\nonumber\\
\nonumber\\
&\theta\dot{\eta}=\dive\Cb(\nex\,,\theta)[\nabla\theta]+r+\delta(\nex\,,\theta\,,\mub\,,\nabla\mub)\,,\mbox{ in }\caR_{t}\,,\nonumber\\
\nonumber\\
&-\epsilon_{o}\Delta\varphi=e\qbels\cdot\nex+q^{*}\,;\nonumber\\
\\
&\Mel(\nex\,,\theta)[\nabla\mub]\cdot\mb=\sels\,, \nonumber\\
\nonumber\\
&\Cb(\nex\,,\theta)[\nabla\theta]\cdot\mb=0\,,\mbox{ on } \partial\caR_{t}\,,\nonumber\\
\nonumber\\
&\nabla\varphi\cdot\mb=0\,,\nonumber
\end{eqnarray}
with initial data:
\begin{equation}\label{inidata}
\nex_{o}(x)=\nex(x\,,0)\,,\quad\theta_{o}(x)=\theta(x\,,0)\,,
\end{equation}
with the \emph{micromechanical dissipation} $\delta=\delta(\nex\,,\theta\,,\mub\,,\nabla\mub)$
\begin{equation}\label{microdissipation}
\delta=\Mel(\nex\,,\theta)[\nabla\mub]\cdot\nabla\mub+\Hel(\nex\,,\theta)\mub\cdot\mub\geq 0\,.
\end{equation}

The coupled system (\ref{evolution1}) generalizes, for  scintillators, the equations obtained into \cite{AGH02} for semiconductors (\emph{vid.\/} Theorem {\bf 6.2}, eqn. (19): \emph{vid. also\/} equations (4.2) of  \cite{MIE11}).

We notice that from equations (\ref{evolution1})$_{3,4}$ we have:
\begin{equation}
\frac{d}{dt}\int_{\caR}\eta=\int_{\caR}\theta^{-1}(\dive\Cb[\nabla\theta]+r+\delta)=\int_{\caR}\theta^{-1}(r+\delta)\,,
\end{equation}
which tells us that whenever there are no external heat sources, $r=0$, the total microentropy is non-decreasing:
\begin{equation}\label{entropicdelta}
\frac{d}{dt}\int_{\caR}\eta=\int_{\caR}\theta^{-1}\delta\geq 0\,;
\end{equation}
furthermore, from (\ref{evolution1})$_{1,2}$ the total micromechanical dissipation can be written in terms of the external sources $(\sels\,,\bels)$ to arrive at:
\begin{equation}
\frac{d}{dt}\int_{\caR}\eta=\int_{\caR}\theta^{-1}(\bels-\dot{\nex})\cdot\mub+\int_{\partial\caR}\theta^{-1}\sels\cdot\mub\geq 0\,.
\end{equation}

If we assume to choose the control volume $\caR$ in such a way that $\sels=\bzero$, then by the means of
(\ref{evolution1})$_{1,4}$  the dissipation (\ref{microdissipation}) can be given a simpler expression:
\begin{equation}\label{simpledelta}
\delta=(\bels-\dot{\nex})\cdot\mub\,.
\end{equation}
As a final consequence we notice that whenever the external supply of electric charge density is stationary, \emph{i.e.\/} $\bels=\bzero$, then
\begin{equation}
\delta=-\mub\cdot\dot{\nex}\geq 0\,.
\end{equation}

The total micromechanical  dissipation $\caD$ on the control volume $\caR$
\begin{equation}
\caD=\int_{\caR}\delta\geq 0\,,
\end{equation}
is equivalent, by (\ref{microdissipation}),  to the twice of the \emph{Conjugate Dissipation Functional}:
\begin{equation}\label{twicediss}
\Psi^{*}(\nex\,,\theta\,,\mub)=\frac{1}{2}\int_{\caR}\Mel(\nex\,,\theta)[\nabla\mub]\cdot\nabla\mub+\Hel(\nex\,,\theta)\mub\cdot\mub\,.
\end{equation}

It is easy to show how, as in \cite{MIE11}, the evolution equation (\ref{evolution1})$_{1,4}$ can be put in an equivalent variational formulation 
\begin{equation}\label{gradflow}
\dot{\nex}=-D\Psi^{*}(\nex\,,\theta\,,\mub)\,,
\end{equation}
where $D$ denotes the Frech\'{e}t derivative; for a comprehensive mathematical treatment of the problem \emph{vid. e.g.\/} \cite{AGS05}.

\section{Reaction diffusion-drift equations for scintillators}

\subsection{Non-isothermal scintillation}

To begin with we assume that the control volume $\caR$ is such that homogeneous Neumann-type condition holds for (\ref{evolution1}), \emph{i.e.\/} $\sels=\bzero$. Moreover we assume no heat supply and stationary, if any, external supply of charge density \emph{i.e.\/} $r=0$ and $\bels=\bzero$.

As a second step we detail the microentropy density; as far as the internal contribution, we assume:
\begin{equation}\label{etau}
\eta_{u}(\theta)=\lambda\log\theta\,,
\end{equation}
where $\lambda>0$ is the \emph{latent heat}. For the electrostatic part instead we assume, as in \cite{MIE11}, a Gibbs entropy:\footnote{In \cite{AGH02} an entropy based on the Fermi level was introduced in place of the one we chosed here for simplicity; in \cite{MIE11} it was discussed when and why to prefer one to the other.}
\begin{equation}\label{etae}
\eta_{e}(\nex)=-k_{B}(\nex\cdot\log(\Nel(\nex)\hat{\cels})-\tr\Nel(\nex))\,,
\end{equation} 
where $k_{B}$ is the Boltzmann constant,
\[
\log(\Nel(\nex)\hat{\cels})\equiv(\log\,\hat{c}_{1}n_{1}\,,\log\,\hat{c}_{2}n_{2}\,,\ldots\log\,\hat{c}_{M}n_{M})\,, 
\]
and $\hat{\cels}\equiv(\hat{c}_{1}\,,\hat{c}_2\,,\ldots\,,\hat{c}_{M})$ are normalizing constants, $[c_{k}]=m^{3}$.

Accordingly, by (\ref{reducedenergy1}),  the Gibbs free-microenergy is the sum of a part which depends on $\nex$ and a part which depends solely on the absolute temperature:
\begin{eqnarray}
\psi(\nex\,,\theta)&=&e\varphi\,\qbels\cdot\nex+\theta\,k_{B}(\nex\cdot\log(\Nel(\nex)\hat{\cels})-\tr\Nel(\nex))\nonumber\\
&+&u(\theta)-\lambda\theta\log\theta\,;
\end{eqnarray}
as a consequence of these assumption, by (\ref{const1})$_{3}$, we get an explicit expression for the scintillation potential:
\begin{equation}\label{mudefinition}
\mub=e\varphi\qbels+\theta\,k_{B}\log(\Nel(\nex)\hat{\cels})\,;
\end{equation}
we notice that relation (\ref{mudefinition}) can be inverted to obtain:
\begin{equation}\label{nex1}
\nex=(\exp\Zbm)\cels\,,
\end{equation}
where $\cels\equiv(c_{1}\,,c_{2}\,,\ldots\,,c_{M})$, $c_{k}=(\hat{c}_{k})^{-1}$ and the $M\times M$ matrix $\Zbm$ is defined as
\begin{equation}
\Zbm\equiv\frac{1}{k_{B}\theta}\diag(\mu_{1}-eq_{1}\varphi\,,\mu_{2}-eq_{2}\varphi\,,\ldots\,,\mu_{M}-eq_{M}\varphi)\,.
\end{equation}

From (\ref{mudefinition}) it is easy to evaluate the gradient of the scintillation potential:
\begin{equation}\label{nablamu}
\nabla\mub=e\qbels\otimes\nabla\varphi+\theta\,k_{B}\Nel(\nex)^{-1}\nabla\nex\,.
\end{equation}

With these results the microstress tensor (\ref{const3})$_{1}$ can be written as
\begin{equation}
\Tbm(\nex\,,\nabla\nex\,,\theta)=e\Mel\qbels\otimes\nabla\varphi+\theta\,k_{B}\Mel\Nel^{-1}\nabla\nex\,;
\end{equation}
if we define the $M\times M$ symmetric and positive semidefinite \emph{carrier mobility matrix}
\begin{equation}
\Sel(\theta)=e\Mel(\nex\,,\theta)\Nel(\nex)^{-1}\,,\quad[\Sel]=m^{2}(V\,sec)^{-1}\,,
\end{equation}
and by the means of the Einstein-Smoluchowski relation the \emph{diffusivity matrix}:\footnote{As it is pointed out into \cite{KD12} the diffusivity, and hence the mobility, can be assumed independent on the excitation density.}
\begin{equation}
\Del(\theta)=\frac{k_{B}\theta}{e}\Sel(\theta)\,,\quad[\Del]=m^{2}sec^{-1}\,,
\end{equation}
then the microstress $\Tbm$ represents an electric current density (normalized with respect to the elementary charge $e$) composed by a diffusive part characterized by $\Del$ and a drift part characterized by $\Sel$:
\begin{equation}\label{microstressRD}
\Tbm(\nex\,,\nabla\nex\,,\theta)=\Del(\theta)\nabla\nex+\Sel(\theta)\Nel(\nex)\qbels\otimes\nabla\varphi\,;
\end{equation}
\emph{vid. e.g.\/} equation (4.3) of \cite{MIE11} or (5.2) of \cite{MIE15}, where a similar result was arrived at for semiconductors by starting from a different approach .

We turn our attention now to the interactive microforce:
\begin{equation}
\kels(\nex\,,\theta\,,\mub)=\Hel(\nex\,,\theta)\mub\,;
\end{equation}
if we assume that:
\begin{equation}
\Hel(\nex\,,\theta)=f(\nex\,,\mub\,,\theta)\,\abels(\theta)\otimes\abels(\theta)\,,
\end{equation}
with the function $f(\nex\,,\mub\,,\theta)$ defined as:
\begin{eqnarray}\label{expansion}
f(\nex\,,\mub\,,\theta)&=&\frac{\abels(\theta)\cdot\nex}{\mub\cdot\ab(\theta)}\sum_{k=0}^{\infty}(\cels_{k}(\theta)\cdot\nex)^{k}\geq 0\,,\nonumber\\
&&\\
\lim_{\nex \to 0}f(\nex\,,\mub\,,\theta)&=&0\,,\nonumber
\end{eqnarray}
then the interactive microforce can be rewritten as:
\begin{equation}\label{reactionterm}
\kels(\nex\,,\theta\,,\mub)=\Kel(\nex\,,\theta)\nex\,,
\end{equation}
with:
\begin{equation}\label{reactionterm2}
\Kel(\nex\,,\theta)=(\sum_{k=0}^{\infty}(\cels_{k}(\theta)\cdot\nex)^{k})\abels(\theta)\otimes\abels(\theta)\,.
\end{equation}

By the means of  (\ref{microstressRD}) and (\ref{reactionterm}), then from (\ref{evolution1})$_{1,4}$ we arrive at a fairly general reaction and diffusion-drift equation for the excitation carrier densities evolution:
\begin{eqnarray}\label{RDfinal}
&\dive(\Del(\theta)\nabla\nex+\Sel(\theta)\Nel(\nex)\qbels\otimes\nabla\varphi)-\Kel(\nex\,,\theta)\nex=\dot{\nex}\,,\mbox{ in }\caR_{t}\nonumber\\
&\\
&\Del(\theta)[\nabla\nex]\mb=\bzero\,,\mbox{ on } \partial\caR_{t}\nonumber
\end{eqnarray}
which recovers and generalizes equation (22) from \cite{VA08} (\emph{vid. also\/} \cite{WGLU11}, \cite{LU15}, \cite{LU17}, \cite{GBD15}). 

Moreover, since by (\ref{etau}), (\ref{etae}) and (\ref{mudefinition}) we have:
\begin{equation}
\dot{\eta}=\lambda\theta^{-1}\dot{\theta}-k_{B}\lels\cdot\dot{\nex}=\theta^{-1}(\lambda\dot{\theta}-\mub\cdot\dot{\nex}+e\varphi\,\qbels\cdot\dot{\nex})\,,
\end{equation}
then we get from (\ref{evolution1})$_{2}$ the non-homogeneous heat equation:
\begin{equation}\label{heatfinal}
\lambda\dot{\theta}=\dive\Cb(\nex\,,\theta)[\nabla\theta]-e\varphi\,\qbels\cdot\dot{\nex}\,,\quad\mbox{ in }\caR_{t}\,,
\end{equation}
with an electrostatic source term $r_{e}=-e\varphi\,\qbels\cdot\dot{\nex}$ (\emph{cf.\/} equation (4.2) of \cite{MIE11}).

The non-isothermal evolution problem for a scintillator is given by  (\ref{RDfinal}),  (\ref{heatfinal}) and (\ref{evolution1})$_3$ with homogeneous Neumann boundary condition the and initial data (\ref{inidata}).

{\remark Charge conservation}

\noindent Scintillation depends on the evolution of charge carriers: accordingly we must require that trough the whole process the electric charge is conserved. Let $Q=Q(t)$ be the \emph{total electric charge} in $\caR$:
\begin{equation}
Q(n)=Q^{*}+\int_{\caR}e\qbels\cdot\nex\,,\quad Q^{*}=\int_{\caR}q^{*}\,,
\end{equation}
then by (\ref{laplace}) with Neumann boundary condition we must have:
\begin{equation}\label{cons}
Q^{*}+\int_{\caR}e\qbels\cdot\nex=0\,,\quad \forall t\in[0\,,\tau)\,.
\end{equation}
We remark that (\ref{cons}) is the necessary condition to have an unique weak solution $\varphi\in H^{1}(\caR)$ to equation (\ref{evolution1})$_{3,6}$ such that $\overline{\varphi}=0$, where 
$\overline{f}$ denotes the mean value of $f$ on $\caR$.

Moreover (\ref{cons}) leads to the conservation law
\begin{equation}\label{chargecons}
\frac{\dif}{\dif t}Q=\int_{\caR}e\qbels\cdot\dot{\nex}=0\,, \quad\forall t\in[0\,,\tau)\,;
\end{equation}
from  (\ref{chargecons}) and (\ref{RDfinal}) with $\bels=\bzero$ and $\sels=\bzero$ then we obtain:
\begin{equation}\label{chargecons1}
\int_{\caR}\Kel(\nex)\nex\cdot\qbels=0\,,\quad\forall t\in[0\,,\tau)\,,
\end{equation}
which in turn, by definition (\ref{reactionterm2}), implies the constitutive restriction on $\abels$:
\begin{equation}\label{stechio}
\qbels\cdot\abels=0\,.
\end{equation}

\hfill$\square$

{\remark Stationary solutions}

\noindent \noindent We say $(\nex_{\infty}\,,\varphi_{\infty}\,,\theta_{\infty})$ a \emph{stationary solution} of (\ref{RDfinal}) if it solves the elliptic problem:
\begin{eqnarray}\label{elliptic}
&\dive\Mel[\nabla\mub_{\infty}]-\Hel\mub_{\infty}=\bzero\,,&\nonumber\\
&&\nonumber\\
&-\epsilon_{o}\Delta\varphi_{\infty}=e\qbels\cdot\nex_{\infty}\,,&\mbox{ in }\caR\,,\\
&&\nonumber\\
&\dive\Cb[\nabla\theta_{\infty}]=-e\varphi_{\infty}\qbels\cdot\nex_{\infty}&\nonumber\,,
\end{eqnarray}
with homogenous Neumann-type  boundary  conditions and where $\mub_{\infty}=\mub_{\infty}(\nex_{\infty}\,,\varphi_{\infty}\,,\theta_{\infty})$. 

Trivially, this problem admits the solution $\mub_{\infty}=\bzero$ which, by (\ref{nex1}), yields:
\begin{eqnarray}\label{nex2}
\nex_{\infty}&=&(\exp\Zbm^{*})\cels\,,\nonumber\\
&&\\
\Zbm^{*}&\equiv&-\frac{e\varphi_{\infty}}{k_{B}\theta_{\infty}}\diag(q_{1}\,,q_{2}\,,\ldots\,,q_{M})\,.\nonumber
\end{eqnarray}

We shall see in the next subsection, as it was also pointed out into \cite{MIE11},  that solutions (\ref{nex2}) are extremely important in order to describe the solution asymptotic behavior, the starting point for the estimate of one of the two important design parameters for scintillating crystals, the Decay Time (the other being the Light Yield).

\hfill$\square$

\subsection{Isothermal scintillation}

\noindent In the isothermal case, with $\theta=\theta_{o}$, from (\ref{reducedmicroen}) and (\ref{entropicdelta}) we get the rate of change of microentropy:
\begin{equation}
\dot{\eta}=-\theta_{o}^{-1}\dot{\varepsilon}\geq 0\,,
\end{equation}
whereas in the evolution equation (\ref{evolution1})$_{1}$ the temperature appears as a parameter in $\Del$ and $\Kel$. 

The boundary problem with initial data, homogenous Neumann-type b.c. and no external current and charge supply:
\begin{eqnarray}\label{RDDmain}
&\dive(\Del\nabla\nex+\Sel\Nel(\nex)\qbels\otimes\nabla\varphi)-\Kel(\nex)\nex=\dot{\nex}\,,\nonumber\\
\\
&-\epsilon_{o}\Delta\varphi=e\qbels\cdot\nex\,,\nonumber
\end{eqnarray}
describes the evolution of excitation carriers in a rigid and defectless crystal: the global existence of \emph{Renormalized} and \emph{Weak solutions} for the boundary value problem (\ref{RDDmain}) was obtained within the context of  chemical reactions and semiconductors in \cite{CHJU17}, \cite{FI17} and extended to scintillators into \cite{DA171}, provided the mobility matrix, the interactive microforce, the electric field, the initial data and initial entropy obey some boundness and regularity hypothesis. We refer to \cite{DA171} for further details and simply remember here that the existence of a weak solution allows for numerical solution of (\ref{RDDmain}) with non-linear finite-element-like methods. 

Equations (\ref{RDDmain}) describe the evolution of excitation carriers in a scintillator by the means of three competing mechanisms, namely the Diffusion, the Drift and the Reaction: to analyze the contribution of each one of them we put must obtain a  dimensionless form for these evolution equation. Hence, let $L$ be a characteristic length and $T$ a characteristic time: then we define the dimensionless quantities:
\begin{equation}
z=L^{-1}x\,,\quad \tau=T^{-1}t\,,
\end{equation}
and the dimensionless variables:
\begin{equation}
\nib=L^{3}\nex\,,\quad\psi=\frac{\epsilon_{o}L}{e}\varphi\,,
\end{equation}
to arrive at, from (\ref{RDDmain}):
\begin{eqnarray}
&\dive_{z}(\Del^{*}\nabla_{z}\nib+\Sel^{*}\Nel(\nib)\qbels\otimes\nabla_{z}\psi)-\Kel^{*}\nib=\nib_{\tau}\,,\nonumber\\
&\\
&-\Delta\psi=\qbels\cdot\nib\nonumber\,,
\end{eqnarray}
where the dimensionless matrices $\Del^{*}$, $\Sel^{*}$ and $\Kel^{*}$ are defined as:
\begin{equation}
\Del^{*}=\frac{T}{L^{2}}\Del\,,\quad\Sel^{*}=\frac{eT}{\varepsilon_{o}L^{3}}\Sel\,,\quad\Kel^{*}=T\Kel\,.
\end{equation}

We finally set
\begin{eqnarray}
\Theta=d^{-1}\Del^{*}\,,\quad d=\|\Del^{*}\|\,,\nonumber\\
\Xi=m^{-1}\Sel^{*}\,,\quad m=\|\Sel^{*}\|\,,\\
\Upsilon=h^{-1}\Kel^{*}\,,\quad h=\|\Kel^{*}(\nex_{o})\|\,,\nonumber
\end{eqnarray}
to arrive at an expression of (\ref{RDDmain}) in terms of the dimensionless parameters $(d\,,m\,,h)$ which accounts respectively for the magnitude of Diffusion, Drift and Reaction
\begin{eqnarray}\label{adimensionale}
&d\,\dive_{z}\Theta\nabla_{z}\nib+m\,\dive_{z}\Xi\Nel(\nib)\qbels\otimes\nabla_{z}\psi-h\,\Upsilon\nib=\nib_{\tau}\,,\nonumber\\
&\\
&-\Delta\psi=\qbels\cdot\nib\nonumber\,.
\end{eqnarray}

For the boundary value problem (\ref{adimensionale}) we obtained into \cite{DA171}, by adapting to scintillators the results obtained in \cite{FK18} for semiconductors, an estimate of the asymptotic decay of solutions. We give here only the main result, the details being presented in full into \cite{DA171}. 

For $(\nib_{\infty}\,,\psi_{\infty})$ a stationary solution of (\ref{adimensionale}) we have:
\begin{equation}
\|\nib-\nib_{\infty}\|^{2}+\|\psi-\psi_{\infty}\|^{2}\leq C_{2}G(\nex_{o}\,,\varphi_{o})e^{-C_{1}t}\,,
\end{equation}
where $G(\nex_{o}\,,\varphi_{o})$ is the total Gibbs free-energy on $\caR$ evaluated for the initial data and the parameters $C_{1}$ and $C_{2}$ depends explicitly on the equation parameters. The value of $C_{1}$ is an estimate for the \emph{Decay time} $\tau_{d}$:
\begin{equation}\label{decaytime}
\tau_{d}=C_{1}^{-1}=\frac{1}{2}e^{2\Phi}\max\{\frac{T}{m}e^{2\Phi}\,,\frac{T}{h_{o}}\}(1+L(\caR))e^{2\Phi}\,,
\end{equation}
where $L(\caR)$ is the Poincar\'e constant of $\caR$, $h_{o}=\|\Kel^{*}(\bzero)\|$ and $\Phi=\|\qbels\psi_{\infty}\|$. As far as we know, (\ref{decaytime}) is the only analytical estimate of the decay time, to date.

Equation (\ref{adimensionale})$_1$ admits many different cases, depending on the relative values of dimensionless parameters; however, two of these cases yield the most popular phenomenological models for scintillator. Indeed, when $h>>\max\{d\,,m\}$, then the reactive term dominates the behavior of the solution and we obtain the so-called \emph{Kinetic} model:
\begin{equation}\label{kinetic1}
-\Kel(\nex)\nex=\dot{\nex}\,;
\end{equation}
when instead it is $d>>\max\{m\,,h\}$ we recover within the present treatment the \emph{Diffusive} model
\begin{equation}\label{diffuse1}
\dive\Del[\nabla\nex]=\dot{\nex}\,.
\end{equation}

\subsection{The Kinetic Model}

One of the most popular phenomenological model for scintillation describes the phenomena in terms of rate equations, like the ones used for the kinetics of chemical reactions \cite{WA57}-\cite{KOKU96}:
\begin{equation}\label{rate}
\dot{\nex}=\gels(\nex)\,,
\end{equation}
and accordingly is aptly-named the \emph{Kinetic Model} (\emph{vid.\/} \cite{BM09} and the references to Chap. 6 of \cite{SW15}): such a model corresponds to the case (\ref{kinetic1}) when $h>>\max\{d\,,m\}$ and provided we identify the reaction term with the interactive microforce:
\begin{equation}
\gels(\nex)=-\kels(\nex)=-\Kel(\nex)\nex\,.
\end{equation}

The term $\gels(\nex)$ is selected to account for three basic kind of non-proportional interaction mechanisms \cite{SW15}: a linear one, which describes the emission or quenching associated with the decay of an emitting center (typically, the energy transfer between one donor and one acceptor); a quadratic, which  describes the emission or quenching associated with the decay of electron-holes pairs and finality a cubic quenching which accounts for the so-called {\em Auger effect} (an electron-hole pair transfers the energy to a third particle which in turns decay by means of a non-radiative process).

This physical behavior corresponds to an interactive microforce which is at most cubic into $\nex$ and which can be recovered if we, as a constitutive assumption, truncate (\ref{expansion}) for $k=2$ to get:
\begin{equation}\label{Kinetics}
\Kel(\nex)=\Ael+\frakA[\nex]+\fourA[\nex\otimes\nex]\,,
\end{equation}
where $\Ael=\abels\otimes\abels$ and the third- and fourth-orders matrices $\frakA$ and $\fourA$ given respectively by:
\begin{equation}
\frakA=\Ael\otimes\cels_{1}\,,\quad\fourA=\Ael\otimes(\cels_{2}\otimes\cels_{2})\,.
\end{equation}

The process of recombination is carefully described into \cite{SW15} and we know that the excitation carriers can recombine by means of both \emph{radiative processes} which are associated to photons production and \emph{non-radiative processes} which instead simply recombine excitation carriers without photons production: further some excitons can convert into electron-hole pairs and vice-versa. Accordingly we can split $\Ael$ and $\frakA$ into:
\begin{equation}\label{radnonrad}
\Ael=\Pel+\Gel+\Eel\,,\quad\frakA=\frakR+\frakG\,.
\end{equation}
The radiative processes are described by  $\Pel=\Pel(\theta)$, which accounts for the short-range radiative processes, namely those associated with the decay of an emitting center (typically, the the energy transfer between one donor and one acceptor, such as exciton-mediated transfer to an activator) and by  $\frakR=\frakR(\theta)$ which describes the quadratic emission associated, for instance, with the decay of $(e-h)$ pairs such as sequential electron and hole capture on an
activator. 

The non-radiative processes are instead described by  $\Gel=\Gel(\theta)$, which describes the short-range quenching  associated with the decay of an emitting center (such as trapping on impurities), by $\frakG=\frakG(\theta)$ which instead describes those associated with the decay of $(e-h)$ pairs (such as the exciton-exciton Auger-like quenching) and by $\fourA=\fourA(\theta)$ which accounts for non-dissipative cubic quenching long-range term (three-body Auger-like quenching such as an $(e-h)$ pair transfers the energy to a third particle which in turns decay by means of a non-radiative process).

The \emph{Exchange matrix} $\Eel=\Eel(\theta)$ finally accounts for the conversion of an excitation carrier into a different kind of particle, like \emph{e.g.\/} the conversion of excitons into electron-hole pairs and vice-versa.

The components of these temperature-dependent parameters can be related to observable phenomena and could be determined experimentally, like it was done in \cite{BM09}, \cite{BM10} and \cite{SI11}; depending on the specific scintillator some of them can be zero in order to describe particular scintillation mechanisms (to this regards see also the discussions in \cite{BM12} and \cite{BM10}). 

As a final remark we observe that the any \emph{homogeneous} stationary solution $\nex_{\infty}$ of (\ref{elliptic}) is also a stationary solution for the corresponding kinetic model.

\subsection{The Diffusive model}

When $d>>\max\{m\,,h\}$, then from (\ref{adimensionale}) we  obtain (\ref{diffuse1}), which recovers the \emph{Diffusive model} \cite{BM12}, \cite{KD12}; however in these models $\Del$ accounts here for the diffusion in the direction \emph{orthogonal} to the track and is used to study the diffusion of excitation carriers within the track in the very first stages of scintillation.

Moreover this model however it is rarely used alone but coupled to a certain degree with some reactive terms like in \cite{KQAD12}, \cite{GLUB12}. We remark that the evolution equations in \cite{VA08}, \cite{WGLU11}-\cite{LU17} contain reactive, diffusive and drift terms and accordingly are particular cases of our Reaction and Diffusion-Drift equation (\ref{RDfinal}).

\subsection{An example from \cite{BM09}}

In \cite{BM09} a Kinetic model was proposed for $M=2$, namely $n_{1}=n_{e-h}$ (electron-hole pairs) and $n_{2}=n_{exc}$ (excitons) in order to identify the kinetic parameters for four different materials (NaI:Tl, BaF$_2$, GSO:Ce and LaCl$_3$:Ce). Equations (1) and
and (2) of \cite{BM09} are obtained from (\ref{kinetic1}) provided the following identification of the linear
\[
\Pel\equiv\left[\begin{array}{cc}R_{1eh} & 0 \\0 & R_{1x}\end{array}\right]\,,\quad\Gel\equiv\left[\begin{array}{cc}K_{1eh} & 0 \\0 & K_{1x}\end{array}\right]\,,\quad\Eel=\left[\bzero\right]\,,
\]
quadratic
\[
\frakR_{111}=R_{2eh}\,,\frakR_{222}=R_{2x}\,,\quad\frakK_{111}=K_{2eh}\,,\frakK_{222}=K_{2x}\,,
\]
and Auger terms 
\[
\fourA_{1111}=K_{3eh}\,,\quad\fourA_{2222}=K_{3x}\,.
\]

Since the electron-hole pairs and the excitons are electrically neutral, then $q_{1}=q_{2}=0$ and accordingly from (\ref{elliptic}) and (\ref{nex2}) with $\theta=\theta_{o}$ we have:
\begin{equation}
\varphi_{\infty}=0\,,\quad \nex_{\infty}=\cels\,.
\end{equation}

The asymptotic solution is homogeneous and hence must solve $\Kel(\cels)\cels=\bzero$: from the calculated values in \cite{BM09} it easy to obtain that for all the four materials studied we must have $\nex_{\infty}=\bzero$. With the data provided then we can evaluate both $h$ and $h_{o}$. However it is more difficult to have reliable estimates of the mobilities and hence of $d$ and $m$. 

For NaI:Tl there are some mobility estimates in \cite{POMU72} and \cite{DIMU72}, however we follow \cite{WGLUBK13}  where for a fixed electrons mobility $\mu_{e}$ a whole range of hole mobilities $\mu_{h}$ was proposed. Since in the model of \cite{BM09} electron and holes are represented by one variable we assume $\mu_{e}=\mu_{h}=\mu_{eh}=\Sel_{11}=8\mbox{ cm}^2/(\mbox{V\,sec})$ with an exciton mobility $\mu_{ex}=\Sel_{22}=0$.

For $L=1$ nm, $T$=1 ns, and $\theta_{o}=293$ K, then we obtain the following values:
\begin{eqnarray*}
d&=&0,02\,,\\
m&=&0,021\,,\\
h(\mbox{1 KeV})&=&0,0815\,,\\
h(\mbox{10 KeV})&=&0,039\,,\\
h(\mbox{100 KeV})&=&0,0081\,,\\
h_{o}&=&0,0066\,.
\end{eqnarray*}

From these values we can observe that at low energy ($E$=1 KeV) $h$ is nearly four times bigger than $m$ and $d$: the diffusion-drift term gives less contribution to the evolution equation than the reactive one and accordingly a purely kinetic model can be used as a good approximation. 

However, for increasing energies the two contributions becomes similar at about $E$=10 KeV and the evolution is described by the full equation (\ref{adimensionale}) whereas for higher energies (about  100 KeV) the diffusion-drift behavior dominates over the kinetic one.

As far as the decay time is concerned, since $\psi_{\infty}=0$ and $L(\caR)=0,068$ for $\caR$ an unit sphere,  we have the estimate:
\begin{equation}
\tau_{d}=C_{1}^{-1}=\frac{1}{2}\max\{\frac{T}{m}\,,\frac{T}{h_{o}}\}(1+L(\caR))=249\,\mbox{ns}\,,
\end{equation}
a result which is consistent with the typical measured values of $230\,$ns \cite{HEIM89} or with recent results of $239\pm3\,$ns given into \cite{SMSC12}.

\subsection*{Acknowledgments}          

This work is within the scope of the CERN R\&D Experiment 18, Crystal Clear Collaboration (CCC) and was supported by the COST Action TD-1401 Fast Advanced Scintillator Timing (FAST). The author wishes to thanks Paul Lecoq and Andrei Vasil'ev for their continuous support toward this line of research. The author wishes also to thank the Referees for their useful remarks and suggestions.


\end{document}